\documentclass[10pt,twocolumn,letterpaper]{article}

\usepackage{iccv}
\usepackage{times}
\usepackage{epsfig}
\usepackage{graphicx}
\usepackage{amsmath}
\usepackage{amssymb}
\usepackage{algorithm}
\usepackage{algorithmic}
\usepackage{bbm}
\usepackage{amsfonts}
\usepackage{mathrsfs}
\usepackage{multirow}
\usepackage{listings}
\usepackage[frozencache=true,cachedir=minted-cache]{minted}

\usepackage[pagebackref=true,breaklinks=true,letterpaper=true,colorlinks,bookmarks=false]{hyperref}

\iccvfinalcopy 

\def\iccvPaperID{406} 

\begin{document}

\title{Retail Market Analysis}

\author{Ke Yuan{$^{1}$}{$^{\dag}$},~ Yaoxin Liu{$^{1}$}{$^{\dag}$},~ Shriyesh Chandra{$^{1}$}{$^{\dag}$},~ Rishav Roy{$^{1}$}{$^{\dag}$}\\\vspace{-8pt}{\small~}\\
{$^{1}$}New York University\\
{\small{{$^{\dag}$}Contribute equally~~ }}\\
{\small\texttt{\{ky2591, yl5739, sc10670, rr4577\}@nyu.edu}}
}

\maketitle
\thispagestyle{empty}

\begin{abstract}
This project focuses on analyzing retail market trends using historical sales data, search trends, and customer reviews. 
By identifying the patterns and trending products, the analysis provides actionable insights for retailers to optimize inventory management and marketing strategies, ultimately enhancing customer satisfaction and maximizing revenue.
\end{abstract}

\section{Introduction}

In today’s competitive retail industry, understanding consumer behavior and market trends is important for success. This project uses data analysis to bridge the gap between what consumers want and what sellers provide. By examining datasets such as Instacart transactions, Amazon reviews, Google Trends data, Percent Change in Consumer Spending, and Walmart sales, the study identifies key patterns and trends influencing purchasing decisions. These insights help retailers make better decisions, ensuring products are available during busy periods, reducing waste, and improving customer satisfaction. This project highlights how analyzing data can lead to more efficient and effective retail operations. 

\section{Method}
We processed and analyzed data collected from various sources, including Amazon Reviews, Walmart, Instacart Market Basket, Percent Change in Spending, and Google Trends, all stored in HDFS (Hadoop Distributed File System). The workflow is divided into three main stages: Storage, Preprocessing, and Analysis/Results.

\textbf{Storage}: Data from diverse sources is stored in HDFS, ensuring scalable and efficient data access for large-scale processing.

\textbf{Preprocessing}: Utilizing Spark SQL, this stage involves:
\begin{itemize}
\item Data Cleaning: Removing inconsistencies and ensuring the data is ready for analysis.
\item Data Profiling: Understanding the structure and quality of the datasets.
\item Data Transformation: Structuring the data into a suitable format, saved as parquet files for optimized storage and access.
\end{itemize}

\textbf{Analysis and Results}: This stage leverages Spark MLlib for advanced analytics. The Normalized Dataset is input to Machine Learning Models to derive predictive insights.
Results are analyzed through Statistics and presented visually using Visualizations.

Figure 1 shows the data flow diagram of the project.

\begin{figure}[h]
    \centering
    \includegraphics[width=\linewidth]{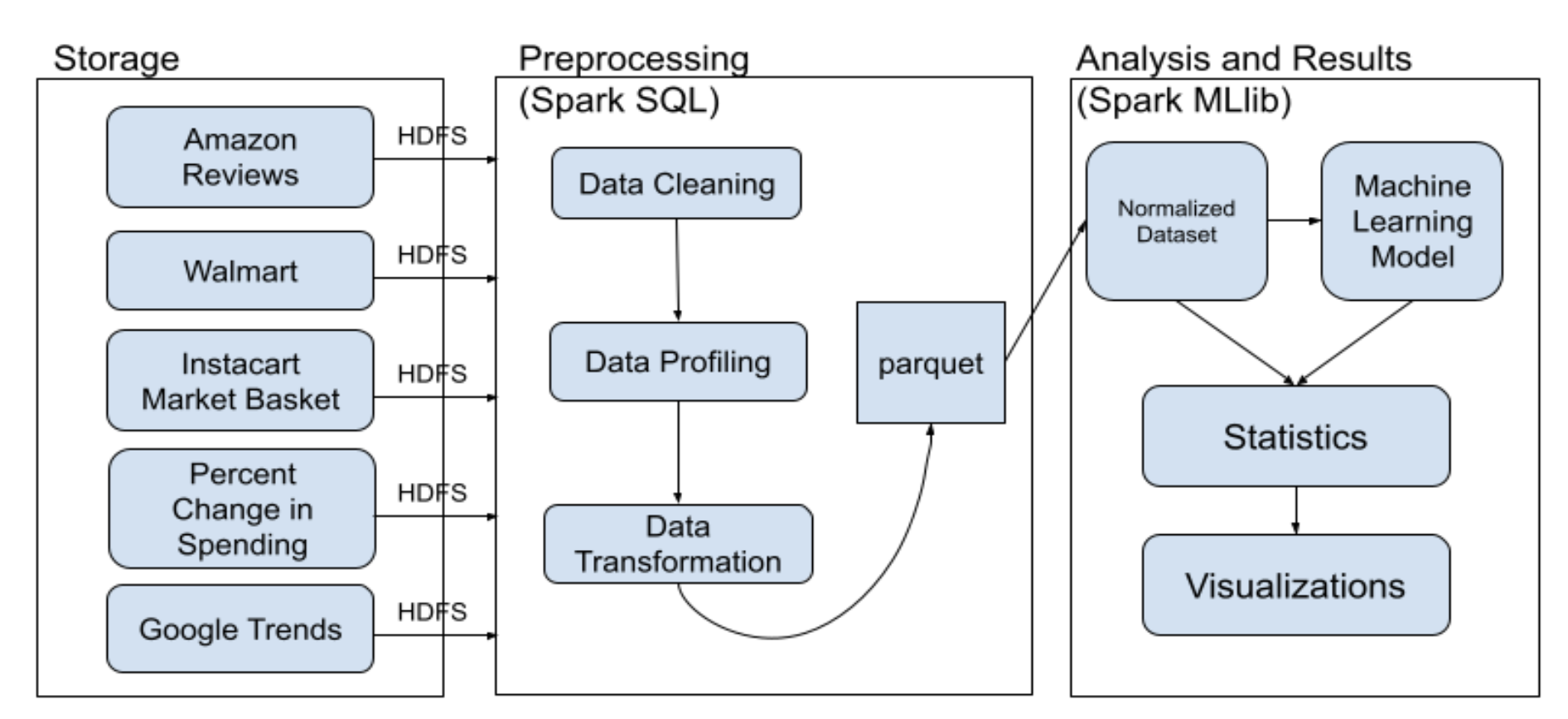}
    \caption{Design Diagram}
    \label{fig:enter-label}
\end{figure}

\section{Google Trends Data}
The cleaned Google Trends dataset contains weekly data for 162 keywords over the past 2 years. These keywords represent broad product categories (e.g., desk, tea, sofa) of commonly consumed everyday items. The trends data are normalized value from 0-100.

\subsection{Feature Engineering}
Currently, the dataset includes only one feature: the date (yyyy-mm-dd). Therefore, additional features need to be extracted before applying a machine learning model to predict trends.

\textbf{Year, Month and Day:}
As a first step, the date is separated into three features: year, month, and day, to better represent time. Each feature is represented as an integer.

\begin{Verbatim}
val newDs = ds
    .withColumn("year", col("date")
        .substr(1, 4).cast("int"))
    .withColumn("month", col("date")
        .substr(6, 2).cast("int"))
    .withColumn("day", col("date")
        .substr(9, 2).cast("int"))
\end{Verbatim}

\textbf{Season: }
Next, a new feature, season, is added based on the month. As a decision tree model requires numerical encoding for each feature, the season feature is manually encoded as 1, 2, 3, and 4 to represent spring, summer, fall, and winter. The reason for not using one-hot encoding is that decision tree models do not require it.

\begin{Verbatim}
def f_season(m: Int): Int = {
        ......
    }

// Register the function as a UDF
val f_season_udf: UserDefinedFunction 
    = udf((m: Int) => f_season(m))

// Add the season feature using the UDF
val newDs2 = newDs
.withColumn("season", f_season_udf(col("month")
.cast("int")))
\end{Verbatim}

\textbf{Is\_holiday: }
Since holidays have a significant effect on consumer decisions, the is\_holiday feature is important. We select several major holidays, including Christmas, New Year, Thanksgiving, Black Friday, and Valentine's Day. If the date is near any of these chosen holidays, the is\_holiday feature is set to 1, otherwise 0

\begin{Verbatim}
// Define a function to mark holidays
val isHoliday = when(
            ...... ,
    1
).otherwise(0)

// Add the holiday feature
val newDs3 = newDs2
.withColumn("is_holiday", isHoliday)
\end{Verbatim}

\textbf{day\_of\_week and is\_weekend: }
The weekend also has a significant effect on consumer decisions because people generally have more time to shop on weekends. Therefore, the is\_weekend feature is included in the model. Additionally, the day\_of\_week feature provides deeper insights and is more precise for analysis, so it is also included in the set of features.

\begin{Verbatim}
val newDs4 = newDs3
.withColumn("day_of_week", 
    dayofweek(col("date")))
.withColumn("is_weekend", 
    when(col("day_of_week") === 1 
    || col("day_of_week") === 7, 1)
.otherwise(0))
\end{Verbatim}

\textbf{sin\_season and cos\_season}
Both features reflect the cyclical nature of seasons. For example, the transition from December (winter) to January (winter) should not be treated as a large gap, but as a smooth and continuous transition. Using sine and cosine functions helps in encoding this continuity and improved Model Performance.

\begin{figure}[h]
    \includegraphics[width=\linewidth]{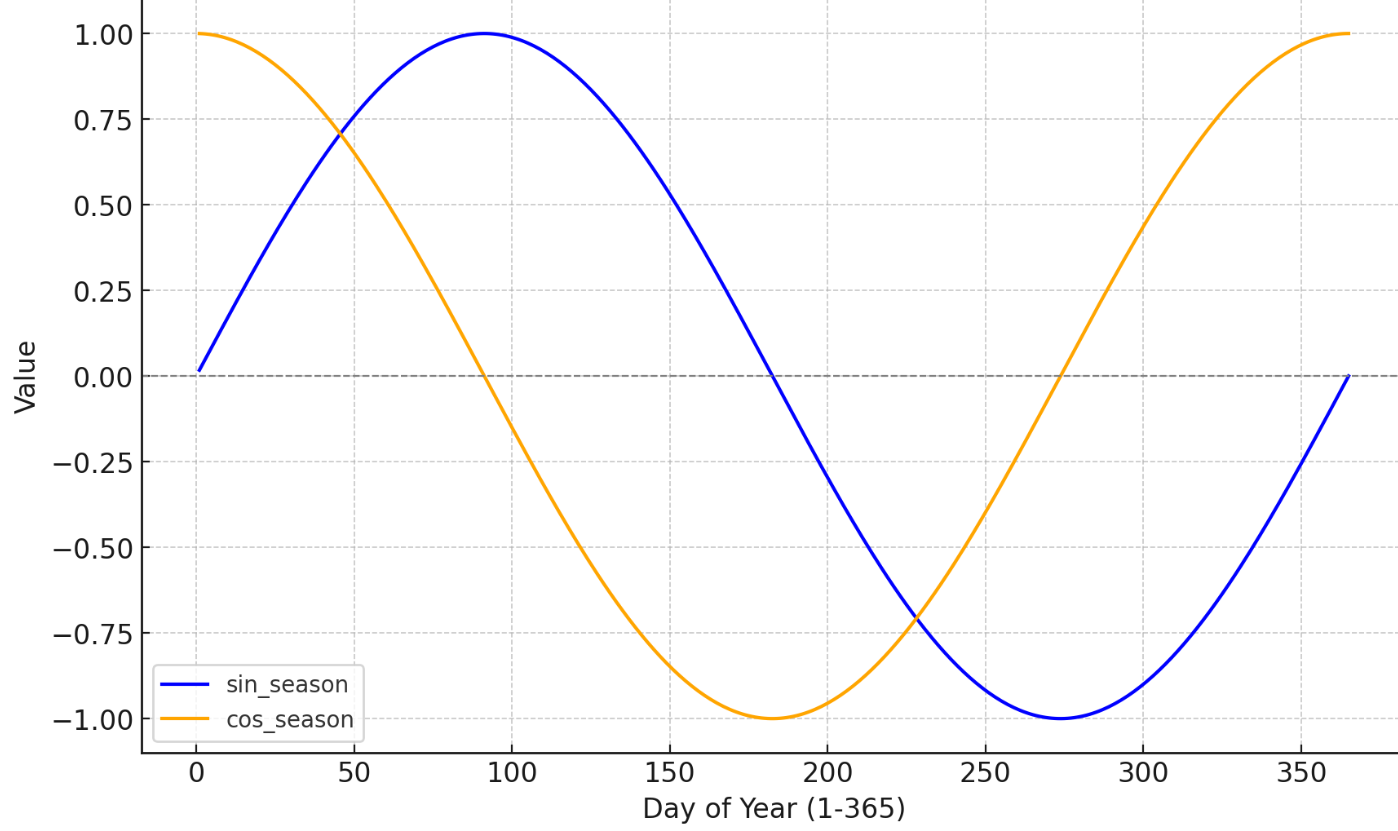}
    \caption{Illustration of sin\_season and cos\_season}
    \label{fig:enter-label}
\end{figure}

\begin{Verbatim}
val newDs5 = newDs4
.withColumn("day_of_year", 
    dayofyear(col("date")))
.withColumn("sin_season", 
    sin(lit(2 * math.Pi) 
    * col("day_of_year") 
    / lit(365)))
.withColumn("cos_season", 
    cos(lit(2 * math.Pi) 
    * col("day_of_year") 
    / lit(365)))
.drop("day_of_year")
\end{Verbatim}

At this stage, we have 9 features for future analysis, which encapsulate key information from the original date feature.

\subsection{Prediction}
In this problem, the objective is to predict the trends of keywords using several time-related features. However, the trends at a specific point in time are not always strongly influenced by past trends, making it challenging for traditional time series models to capture the relationship. As a result, a decision tree model is selected, as it can effectively handle the complexity of non-linear relationships and is less dependent on the sequential nature of past data compared to other time series-based machine learning models.
$$
RMSE=\sqrt{\frac{1}{n}\sum_{i=1}^n\left(y_i-\hat{y}_i\right)^2}
$$

The dataset is split into a training set and a testing set, with 80\% used for training and 20\% for testing. RMSE is employed as the evaluation metric.

\begin{Verbatim}
val Array(trainingData, testData) = 
data.randomSplit(
    Array(0.8, 0.2), seed = 1234L
)
\end{Verbatim}

An independent machine learning model was trained for each keyword to make predictions, with each model operating independently. This approach was chosen because there is no significant or direct correlation between different keywords. If a single model were used to perform multi-target predictions for all keywords, it could lead to challenges such as increased model complexity, interference between targets, and ultimately suboptimal prediction performance. By training separate models, we aim to better capture the unique patterns and characteristics of each keyword, improving the overall accuracy and reliability of the predictions.

\begin{Verbatim}
// Train and evaluate the model 
// for each keyword label
keywordLabels.foreach { label =>
    val dt = new DecisionTreeRegressor()
    .setLabelCol(label)
    .setFeaturesCol("features")

    // Train the model
    val model = dt.fit(trainingData)

    // Make predictions
    val predictions = model
        .transform(testData)

    // Evaluate the model using RMSE
    val evaluator = ...
    .setLabelCol(label)
    .setPredictionCol("prediction")
    .setMetricName("rmse")

    ......

    }
\end{Verbatim}

\subsection{Result}

For the results, we calculate the RMSE on the test set and the predicted value for each keyword on 2024-12-31. Below is a portion of the results:

\begin{table}[!ht]
    \centering
    \caption{Part of results}
    \begin{tabular}{|c|c|c|}
    \hline
        \textbf{Keywords} & \textbf{RMSE} & \textbf{Prediction on 2024-12-31} \\ \hline
        Winter Coat & 2.2654 & 10.0 \\ \hline
        Swimsuit & 2.6990 & 16.0 \\ \hline
        Thanksgiving Turkey & 21.7228 & 65.0 \\ \hline
        Easter Eggs & 20.7649 & 6.1071 \\ \hline
    \end{tabular}
\end{table}

The full set of results can be found  \href{https://4j7tqfe2xjelji6rpk2uhyzisq-dot-us-central1.dataproc.googleusercontent.com/gateway/default/jobhistory/joblogs/nyu-dataproc-w-0.c.hpc-dataproc-19b8.internal:8026/container_1724767128407_11524_01_000001/container_1724767128407_11524_01_000001/ky2591_nyu_edu/stdout/?end.time=9223372036854775807&start=0&start.time=0}{here}.

A comprehensive analysis of the RMSE values for all predictions indicates that the predictive model demonstrates a high level of accuracy in this task. The average RMSE across all test cases is 2.6561, demostrating the model's reliability and precision in capturing the underlying patterns in the data.

\begin{table}[!ht]
    \centering
    \caption{RMSE}
    \begin{tabular}{|c|c|c|c|}
    \hline
        \textbf{Min} & \textbf{Max} & \textbf{Mean} & \textbf{Variance} \\ \hline
        0.0 & 21.7228 & 2.6561 & 11.2749 \\ \hline
    \end{tabular}
\end{table}

Keywords with RMSE less than 3.0 and predicted trends value greater than 80.0 are selected for future analysis: \textbf{soap} (trends = 84), \textbf{shampoo} (trends = 92.43), \textbf{camera} (trends = 92), \textbf{smartphone} (trends = 92.9), \textbf{vacuum} (trends = 91), \textbf{microwave} (trends = 90.9) and \textbf{laptop} (trends = 90.1). Figure 3 shows the distribution of predicted values on 2024-12-31.

\begin{figure}[h]
    \includegraphics[width=\linewidth]{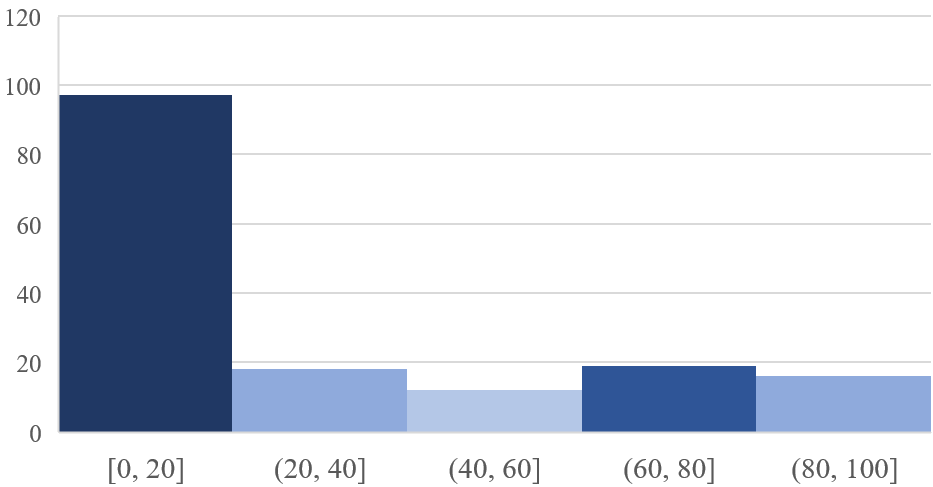}
    \caption{Distribution of Predicted Values}
    \label{fig:enter-label}
\end{figure}

\section{Instacart Market Basket Analysis}

Instacart, a grocery order and delivery app with stores in the United States and Canada, provides a user experience where customers get product recommendations based on their previous orders. Instacart provided transactional data on customer orders over time.

The Instacart data is split and normalized into 6 datasets, each containing key information on orders and products purchased.

\begin{enumerate}
    \item \textbf{aisles.csv} - details of aisles
    \item \textbf{department.csv} - Details of Department
    \item \textbf{products.csv} - details of a product. Each product is associated with an aisle and a department
    \item \textbf{orders.csv} consists of order details placed by any user
    \item \textbf{order\textunderscore products\textunderscore\textunderscore prior.csv} - consists of all product details for any prior order, that is, the order history of every user
    \item \textbf{order\textunderscore products\textunderscore \textunderscore train.csv} - consists of all product details for a train order, that is, current order data for every user. These data contain only 1 order per user.
\end{enumerate}

\subsection{Data Profiling}

Using Spark SQL and the Zeppelin notebook in NYU Dataproc Cluster, we learn the following information from our dataset.
\begin{enumerate}
    \item \textbf{Dataset Size}: There are 134 aisles, 21 departments, 50k products, 3.4 million unique orders, 32 million records on order-products history and 1 million unique training records
    \item \textbf{Null Checks/Missing Values}: None of the dataset has null values, except orders.csv - Orders Dataframe is expected to have null values in the \textbf{days\textunderscore since\textunderscore prior\textunderscore order} column, in case there is no previous order.
\end{enumerate}

\subsection{Data Pre-processing}

The data from the different data frames are merged and joined to convert into a format that would be convenient for further analysis.

These are the pre-processing steps taken:
\begin{enumerate}
    \item Join the smaller tables (aisles, departments and products) to create the productDetails Dataset.
    
    \item Merge the data from \textbf{order\textunderscore products\textunderscore\textunderscore train.csv} and \textbf{order\textunderscore products\textunderscore\textunderscore prior.csv}
    
    \item Orders table has pre-existing null values in the \textbf{days\textunderscore since\textunderscore prior\textunderscore order} column. These can be filled with -1.

    \item Join the Orders and ProductDetails dataset with the merged training dataset to create a single de-normalized dataset.

    \item Remove any null records created from all the joins

    \item Remove all the ID columns and the eval\textunderscore set column from the final dataset
\end{enumerate}

The resultant denormalized dataset is saved in parquet format for future analysis.

\subsection{Market Basket Analysis}

The cleaned and denormalized data set has the following schema:
\begin{figure}[h]
    \includegraphics[width=\linewidth]{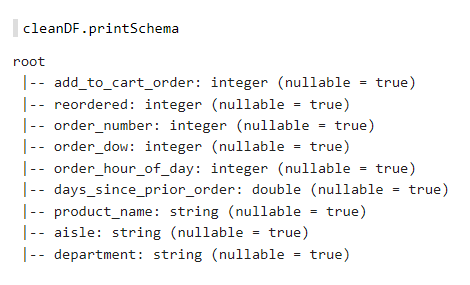}
    \caption{Denormalized Schema for Instacart Data}
    \label{fig:enter-label}
\end{figure}

We perform data analysis on this cleaned data to identify interesting patterns that would prove to be useful to the management team at Instacart to fine-tune their product placements and employee work schedules.

\subsubsection{Orders vs Hours of the Day}

We count the total number of orders for each hour of the day to estimate the number of employees required per shift and ensure efficient staffing. Analysis of the bar graph lets us know that the distribution follows a traditional bell curve with minimal activity during from 12 am to 6 am. Orders increase during the morning and afternoon hours and gradually decrease as it gets late.

\begin{figure}[h]
    \centering
    \includegraphics[width=\linewidth]{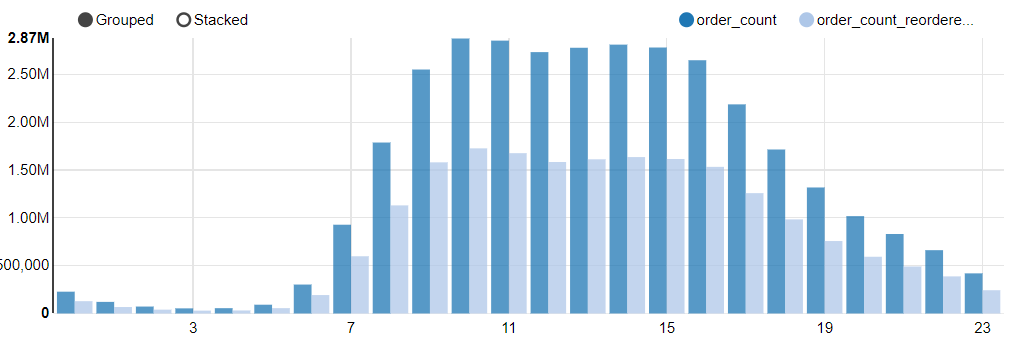}
    \caption{Orders vs Hours of the Day}
    \label{fig:enter-label}
\end{figure}

\subsubsection{Frequently ordered product departments and product aisles}

We investigate the relationship between the number of items ordered per department/aisle to identify the most profitable product  departments/aisles. The analysis has been visualized as a pie chart indicating the percentage of total orders per department/aisle. Since there are over 100 aisles, we pick the top 25 aisles for the pie chart graphic. 

\textbf{Produce and dairy eggs} are the most common departments. \textbf{Fresh fruits and vegetables} are the most common product aisles.

\begin{figure}[h]
    \centering
    \includegraphics[width=\linewidth]{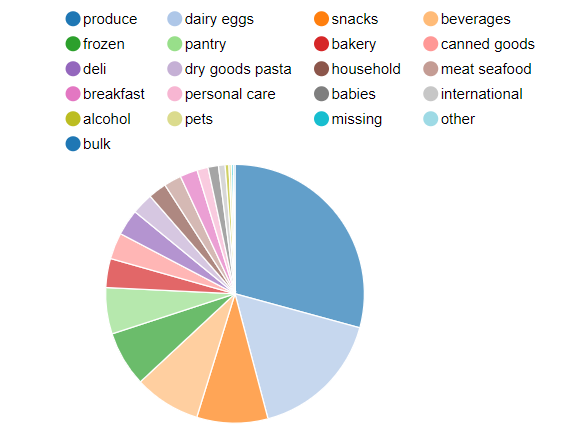}
    \caption{Popular Departments}
    \label{fig:enter-label}
\end{figure}

\begin{figure}[h]
    \centering
    \includegraphics[width=\linewidth]{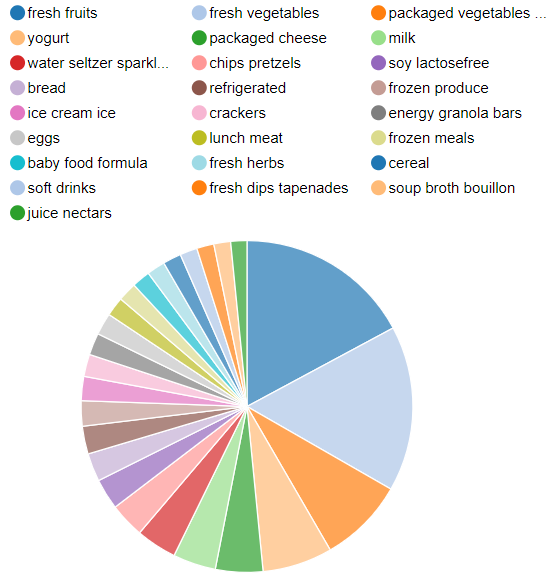}
    \caption{Popular Aisles}
    \label{fig:enter-label}
\end{figure}

\subsubsection{Relationship between position in shopping cart and reorder rate}

There is quite an interesting correlation between the item’s position in the shopping cart and its reorder rate. Reorder rate can be defined as:

\[\textbf{Reorder Rate} = \frac{\textbf{Sum of reordered items}}{\textbf{Total number of items}}\]

From positions 1 through 50, there is a clear correlation between the reorder rate and the position in the shopping cart.

\begin{figure}
    \centering
    \includegraphics[width=1\linewidth]{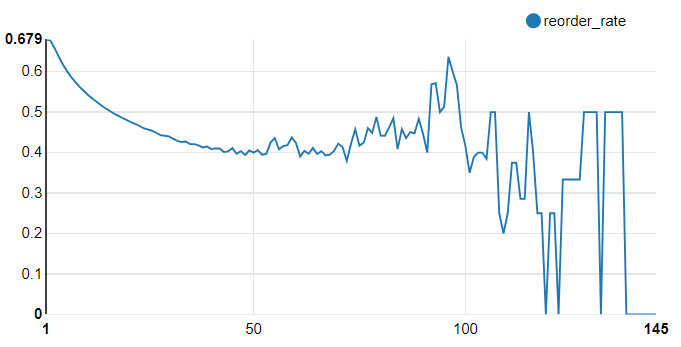}
    \caption{Position in cart vs Reorder rate}
    \label{fig:enter-label}
\end{figure}

\begin{itemize}
    \item \textbf{Early positions}: Higher reorder rates may indicate staple or frequently purchased items.
    \item \textbf{Later positions}: Lower reorder rates may suggest less critical or experimental purchases.
\end{itemize}

\section{Percent Change in Consumer Spending}
This dataset measures daily percentage changes in consumer spending across multiple categories, including groceries, health, entertainment, and transportation, over the years 2020 to 2024.

\subsection{Overview of the dataset}
This dataset includes fields that provide high-level insights into consumer spending patterns, such as percentage changes in categories like food service spending, entertainment spending, grocery spending, and more. Table 3 shows a snippet of the dataset.

\begin{table}[!ht]
    \centering
    \caption{Part of Percent Change in Consumer Spending Dataset}
    \begin{tabular}{|c|c|c|c|}
    \hline
        \textbf{StateCode} & \textbf{Date} & \textbf{AllSpending} & \textbf{...} \\ \hline
         900 & 2020-01-13 & -2.3 & ... \\ \hline
        2500 & 2020-01-13 &  -0.218 & ... \\ \hline
        ... & ... & ... & ... \\ \hline
    \end{tabular}
\end{table}

\subsection{Data Processing}
During the processing of this dataset, we convert daily data from multiple areas daily and calculating a single percentage seasonally and yearly to reflect overall trends at same time. This approach ensures that we maintain both accuracy and proportionality.

\begin{verbatim}
  val yearlyMaxAvg = yearlyAverages
  .withColumn(
    "MaxAvgPercent",
    greatest(
      col("YearAvgAllSpending"),
      col("YearAvgFoodService"),
      col("YearAvgEntertainment"),
      col("YearAvgMerchandise"),
      col("YearAvgGrocery"),
      col("YearAvgHealth"),
      col("YearAvgTransport"),
      col("YearAvgRetailIncGrocery"),
      col("YearAvgRetailExGrocery")
    
  
\end{verbatim}

\begin{figure}[h]
    \includegraphics[width=\linewidth]{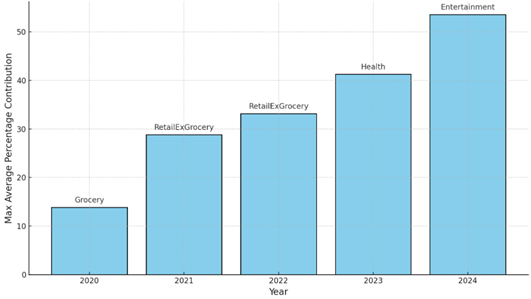}
    \caption{Analysis of percentage change Trends}
    \label{fig:enter-label}
\end{figure}

\subsection{Analysis of yearly and Economic Trends}
Consumer spending patterns vary significantly across seasons and economic conditions. During crises like the COVID-19 pandemic, demand shifted heavily toward essentials such as groceries, reflecting the focus on basic needs. In contrast, stable economic periods saw increased spending on discretionary categories like health, wellness, and entertainment, driven by greater consumer confidence and disposable income.

Figure 9 shows the yearly winner categories by average percentage contribution.

\section{Walmart Retail Dataset}

The Walmart Retail Dataset, obtained from data.world, spans retail order details from 2019 to 2023 and is approximately 300 MB in size. It includes comprehensive fields such as customer demographics, product details, and transaction information, encompassing cities across multiple U.S. states. The dataset served as a foundation for analytical tasks and data exploration, with Zeppelin utilized to perform functions like data transformation, visualization, and aggregation to derive actionable insights. In its initial state the Dataset had a lot many columns that were not necessary at all in our evaluations. Below provides you a picture of the schema of our Dataframe in its inital stage without any preprocessing being performed:

\begin{figure}[h]
    \centering
    \includegraphics[width=0.75\linewidth]{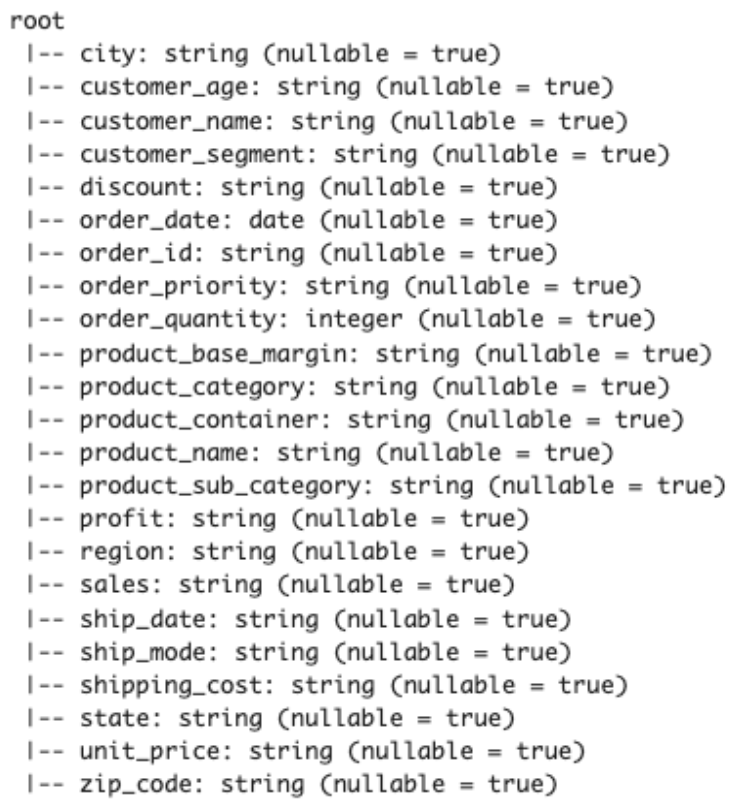}
    \caption{Schema for our original Walmart Dataset before PreProcessing Data}
    \label{fig:Walmart Data without profiling}
\end{figure}

\subsection{Profiling of Data}
Using Spark SQL and the Zeppelin notebook in NYU Dataproc Cluster, we learn the following information from our dataset:

\begin{itemize}
    \item \textbf{Schema and Column Stats:} Identified column types (e.g., strings, dates, integers) and counted null vs. non-null values to assess data quality. Discovered columns like \texttt{profit}, \texttt{sales}, and \texttt{shipping\_cost} contained missing data that needed fixing.
    \item \textbf{Invalid Entries:} Noticed around 18k records in the \texttt{state} column with invalid or malformed data (e.g., numeric or special characters), which required cleaning.
    \item \textbf{Unique Values:} Found 45 unique states and 18 distinct product subcategories. One subcategory was labeled as an unknown category, indicating possible data entry errors or incomplete records.
    \item \textbf{Unnecessary Columns:} Identified several columns that were not required for the analysis, such as \texttt{customer\_name}, \texttt{order\_id}, and \texttt{product\_container}, which were marked for removal.
\end{itemize}

\subsection{Pre-Processing of Data}
\begin{itemize}
    \item \textbf{Data Cleaning:} Replaced or removed null and invalid values. For example, missing values in \texttt{profit}, \texttt{sales}, and \texttt{shipping\_cost} were set to zero. 
    \item \textbf{Filtering Invalid States:} Used regex to filter out invalid state entries (e.g., those containing digits). After dropping these rows, the dataset’s record count reduced from approximately 1,030,000 to 1,016,102.
    \item \textbf{Schema Refinement:} Verified and updated column data types. Automatic schema inference assigned correct types to columns such as dates, integers (e.g., \texttt{order\_quantity}), and strings (e.g., \texttt{city}, \texttt{state}).
    \item \textbf{Removal of Non-Essential Columns:} Dropped columns that were not essential for either the overall analysis or collaborative insights, such as \texttt{order\_priority}, \texttt{product\_base\_margin}, and \texttt{unit\_price}.
\end{itemize}

The resultant Processed dataset with only the essential columns was saved in parquet format for future analysis.

\subsection{Analysis of Walmart Retail Dataset}

\subsubsection{Statewise Analysis of Sales}
From the aggregated results, \textbf{California} and \textbf{Texas} lead in total order quantity, closely followed by \textbf{MA}, \textbf{New Jersey}, and \textbf{Florida}. These top performers also generate correspondingly high profits, suggesting that they represent substantial market share and sales volume within the dataset.

\begin{figure}[H]
    \centering
    \includegraphics[width=\linewidth]{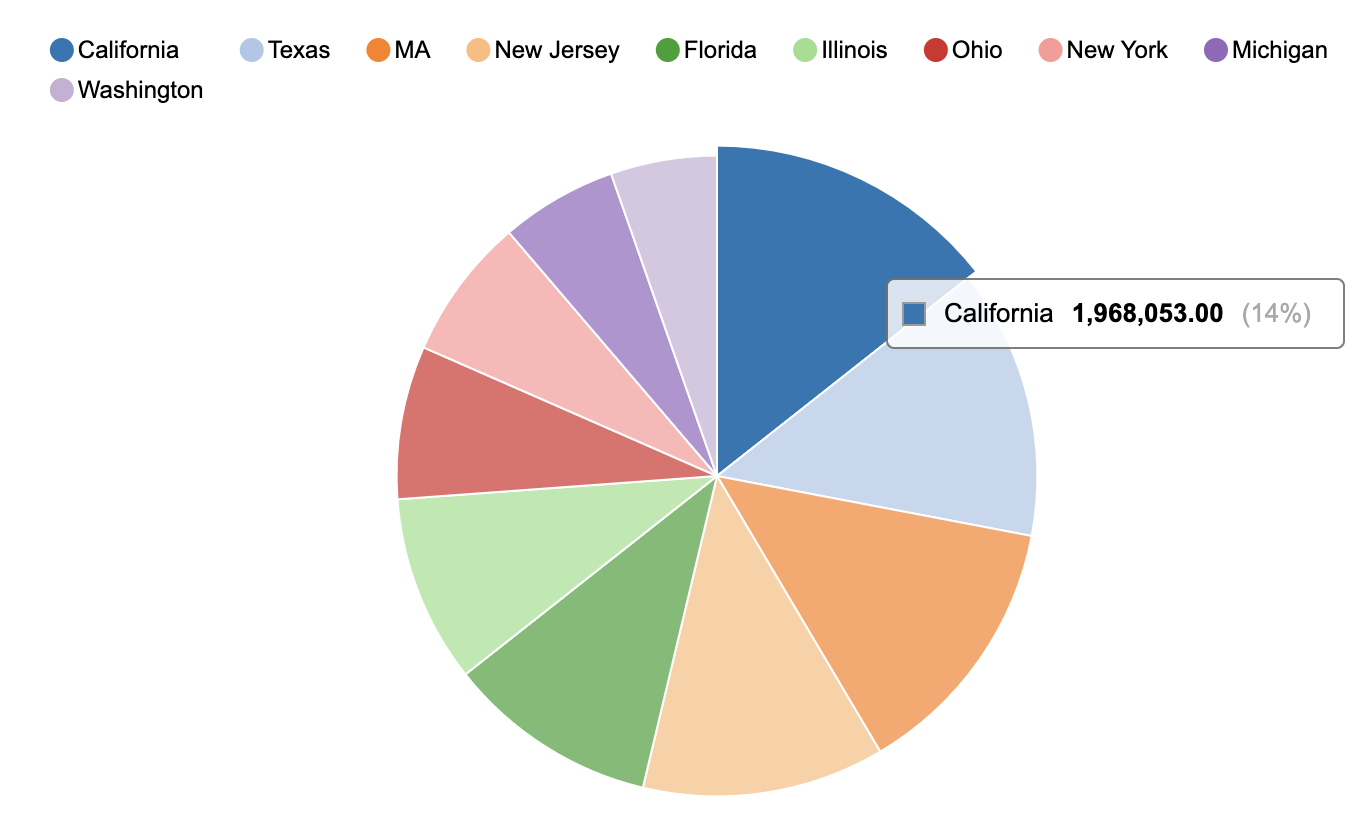} 
    \caption{Statewise Analysis of Sales}
    \label{fig:statewise_sales}
\end{figure}

\subsubsection{Region-Wise Profit Distribution}
When grouping by region (East, West, South, Central), the \textbf{East} emerges as the highest in total order quantity—over \textbf{8.2 million}—and similarly leads in overall profit. The \textbf{South} and \textbf{Central} regions follow behind, indicating potential regional demand differences and logistical variations that might affect profitability. Here we see that even though according to the State-wise distribution California has the most number of orders, the most profit is observed from the East Region while West Region is around half the profit collected from East. We can conclude that even though the number of orders and number of stores is a lot in California, its contribution to the overall West region does not help with the non presence of Walmart in the West.

\begin{figure}[H]
    \centering
    \includegraphics[width=\linewidth]{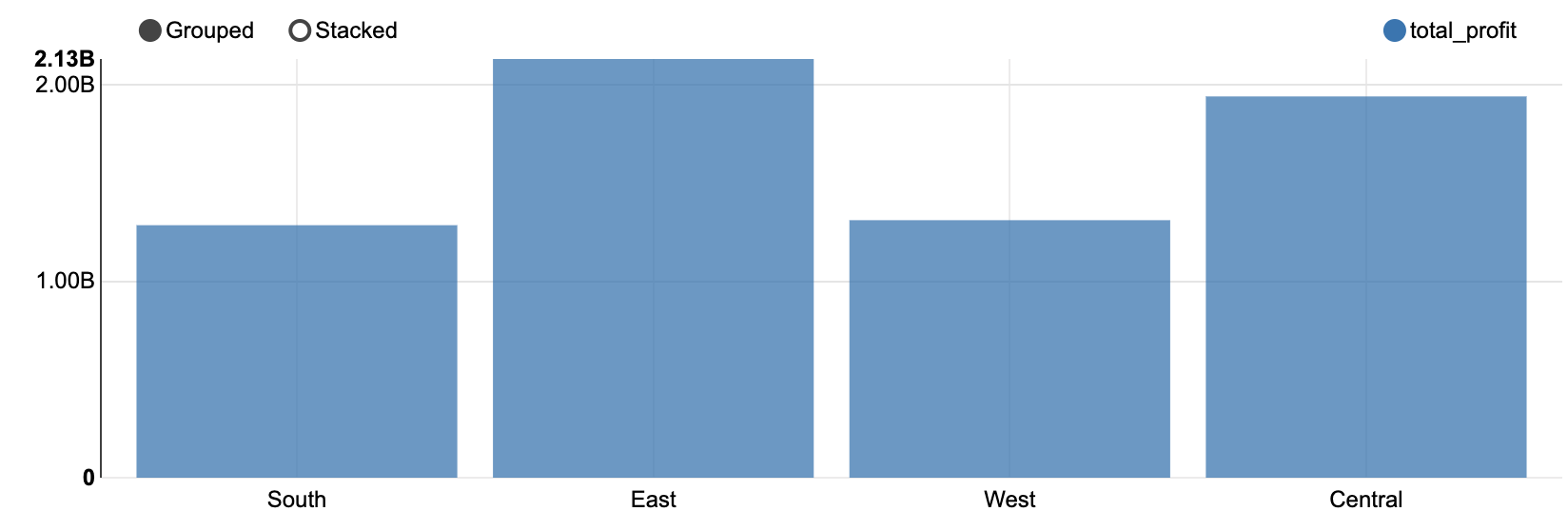} 
    \caption{Region-Wise Profit Distribution}
    \label{fig:regionwise_profit}
\end{figure}

\subsubsection{Number of Cities per State}
States like \textbf{California} (108 unique cities), \textbf{Texas} (102), and \textbf{MA} (101) have the broadest coverage, reinforcing their leading positions in terms of total orders. The large number of cities suggests a widespread presence of Walmart stores and, in turn, a more extensive customer base.

\begin{figure}[H]
    \centering
    \includegraphics[width=1.2\linewidth]{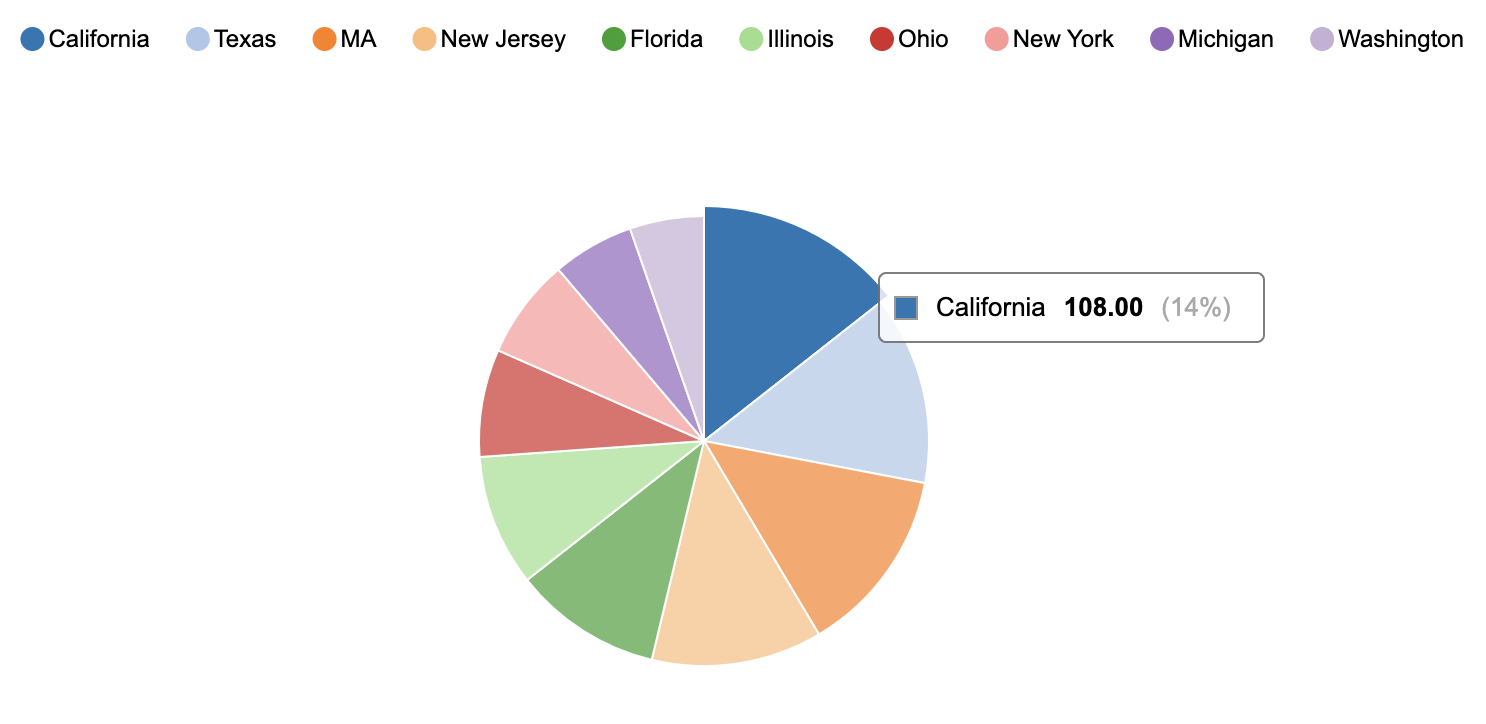} 
    \caption{Number of Cities per State}
    \label{fig:cities_per_state}
\end{figure}

\subsubsection{Customer Age Group Analysis}
Surprisingly, the \textbf{60+} age group accounts for the highest order quantity (over \textbf{10.9 million}) and the greatest total profit. The \textbf{21–40} and \textbf{41–60} brackets are fairly close in their total order volumes (\textasciitilde7.4 million each), while the \textbf{0–20} segment is notably smaller, pointing to specific demographic trends in purchasing behavior. We can realise from this analysis that the shoppers belonging to a younger demographic prefer to not use Commerce brands in person, unlike people of higher age, who have a high presence at in person stores compared to online shopping and E-Commerce.

\begin{figure}[H]
    \centering
    \begin{subfigure}[t]{\linewidth}
        \centering
        \includegraphics[width=\linewidth]{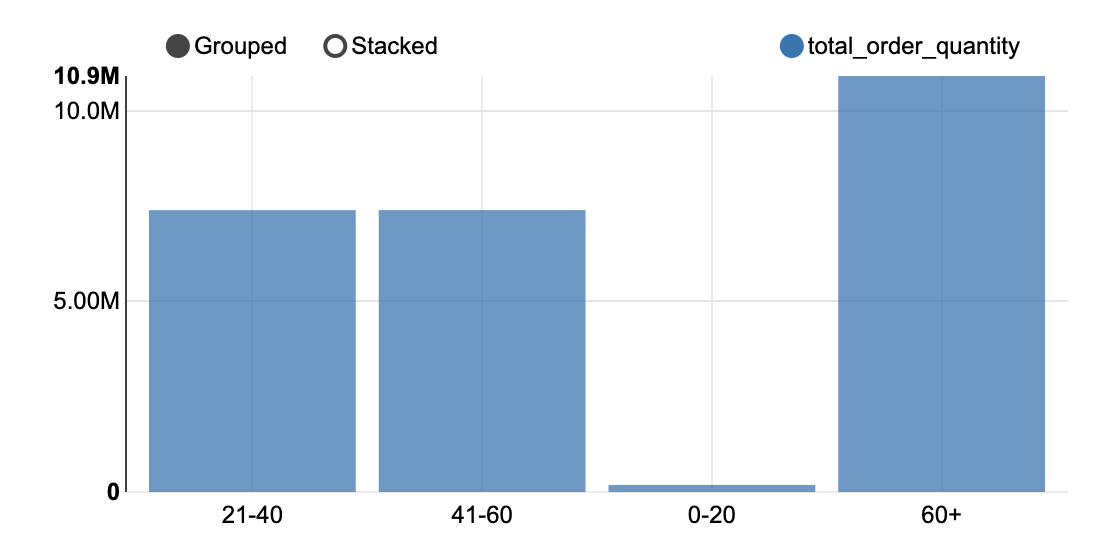} 
        \caption{Customer Age Group Analysis}
        \label{fig:age_group_analysis}
    \end{subfigure}
    \hfill
    \begin{subfigure}[t]{\linewidth}
        \centering
        \includegraphics[width=\linewidth]{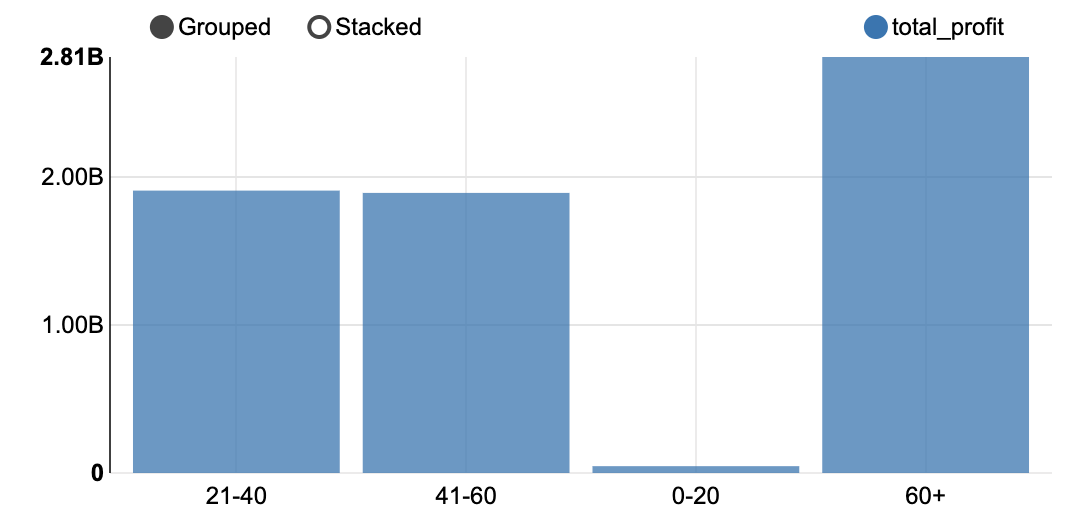} 
        \caption{Analysis of Profits based on Age Group}
        \label{fig:age_group_profits}
    \end{subfigure}
    \caption{Analysis of Order Quantity and Profits Based on Age Group}
    \label{fig:age_group_combined}
\end{figure}
\subsubsection{Monthly Analysis of Order Quantity}
Monthly ordering patterns show a peak in \textbf{January} (around \textbf{2.34 million} orders), followed by some variability throughout the year. There’s a dip in February, and moderate fluctuations continue in subsequent months. This points to potential seasonality and holiday-driven purchasing trends in Q1 and Q4.

\begin{figure}[H]
    \centering
    \includegraphics[width=\linewidth]{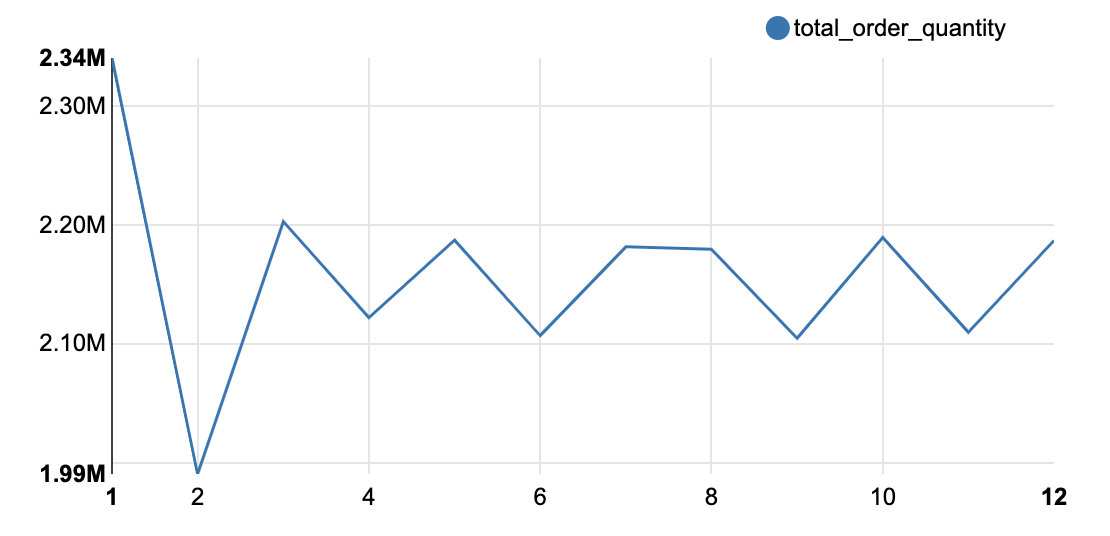} 
    \caption{Monthly Analysis of Order Quantity}
    \label{fig:monthly_analysis}
\end{figure}

\section{Amazon Reviews Dataset}

The Amazon Customer Reviews is a massive dataset (over 75GB) of amazon product reviews.

\smallskip
Typically it has been used for pre-training Large Language Models and in sentiment analysis. We will look at the Pure IDs (5-Core) segment of the dataset which just focuses on the product ratings (out of 5).

\smallskip
A parent ASIN is a unique Amazon Standard Identification Number (ASIN) that groups together products with variations, such as different colors or sizes. The mapping from ASIN to product is
obtained from a large json file (1.25 GB)

\medskip
\noindent
\textbf{Data Links:}
\begin{enumerate}
    \item \href{https://amazon-reviews-2023.github.io/data_processing/5core.html}{Pure IDs (5-Core) - Amazon Reviews'23 }
    \item \href{https://huggingface.co/datasets/McAuley-Lab/Amazon-Reviews-2023/blob/main/asin2category.json}{Hugging Face Mapping Link}
\end{enumerate}

\medskip
In total there are 28 categories of data, each stored in its own zip file. The dataset includes user ratings, and product
metadata for categories, that is the parent ASIN key and user\textunderscore id along with the review timestamp.

\subsection{Data Profiling}
Using Spark SQL and Zeppelin Notebook on NYU Dataproc Cluster, we learn the following information from our dataset.

\begin{enumerate}
    \item \textbf{Dataset Size}: There are approximately 67 million records in the ratings dataset and
approximately 35 million records in the asin dataset.
    \item \textbf{Null Checks/Missing Values}: None of the datasets have any null values
    \item \textbf{Data Summary}: Using the Dataframe API, calculate simple statistics for the Ratings Dataframe. Here are some stats:
    \begin{enumerate}
        \item Ratings range from 1.0 to 5.0
        \item 50\(\%\) of the orders in the dataset have a 5.0 rating
        \item Timestamps are in the Unix timestamp format.
    \end{enumerate}
    \item \textbf{Unknown Category}: When analysing our dataframe after the join we found that unknown was a category with an ASIN of its own.
\end{enumerate}

\subsection{Data Pre-processing}

The data from the different data-frames are merged and joined to get it into a format that would be convenient for further analysis.

\smallskip
There is one pre-processing step that was done while loading the data. Since we load all the ratings csv files into one single dataframe, it helps to create a column called `\textbf{source\textunderscore file}' which refers to the source of the Dataframe Row.

\smallskip
These are the preprocessing steps taken:

\begin{enumerate}
    \item Join the Ratings and ASIN dataset to create a single denormalized dataset.
    \item Replace the nulls generated by the ``left outer join" between the two datasets. We only have null values in the `category' column which we can replace using the value from the `source\textunderscore file' column
    \item Remove the user\textunderscore id, source\textunderscore file and the parent\textunderscore as in column
    \item Convert the UNIX timestamp to a Date format (YYYY-MM-DD) to get a consistent result that can be used to compare with other datasets as well.
    \item Handling the unknown category: This category seemed pretty inconclusive for our analysis and thus we planned on removing it entirely from our final combined dataframe.
    \item The resultant dataset was saved in parquet format for future analysis
\end{enumerate}

\subsection{Data Analysis}

The cleaned and merged dataset has the following structure:

\begin{figure}[h]
    \centering
    \includegraphics[width=0.75\linewidth]{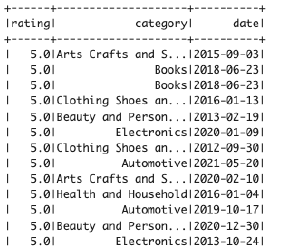}
    \caption{Sample Data from the cleaned Amazon Dataset}
    \label{fig:enter-label}
\end{figure}

We perform data analysis on this cleaned data to identify interesting patterns that would prove to be useful to
the management team at Amazon to fine-tune their inventory management schedules.

\subsubsection{Ratings Count}
We try to look at the ratings count for all the products at Amazon. We notice that most of the products are reviewed at 5 stars (100 M). All other ratings put together have a count of around 50 M. This tells us that review ratings (out of 5) are not a good indicator of analysis on the Amazon products due to their skewed nature.

\begin{figure}
    \centering
    \includegraphics[width=1\linewidth]{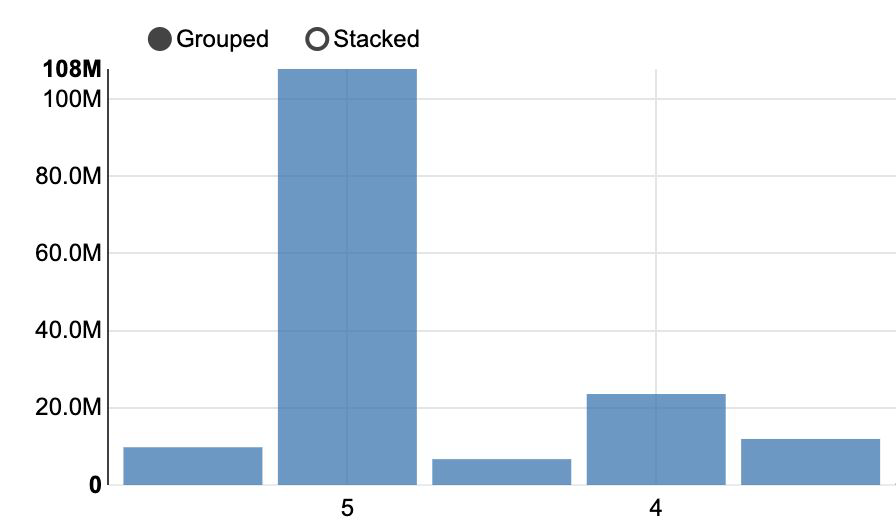}
    \caption{Ratings vs Ratings count}
    \label{fig:enter-label}
\end{figure}

\subsubsection{Category Count}
Next, we look at the distribution of ratings across the different categories in the Amazon Dataset. Here we notice that clothing, shoes, jewelry and electronics are among the top most rated categories. A direct correlation can be drawn to the fact that customers mainly purchase clothing items from Amazon in comparison to Health and Household items.

\begin{figure}[h]
    \centering
    \includegraphics[width=1\linewidth]{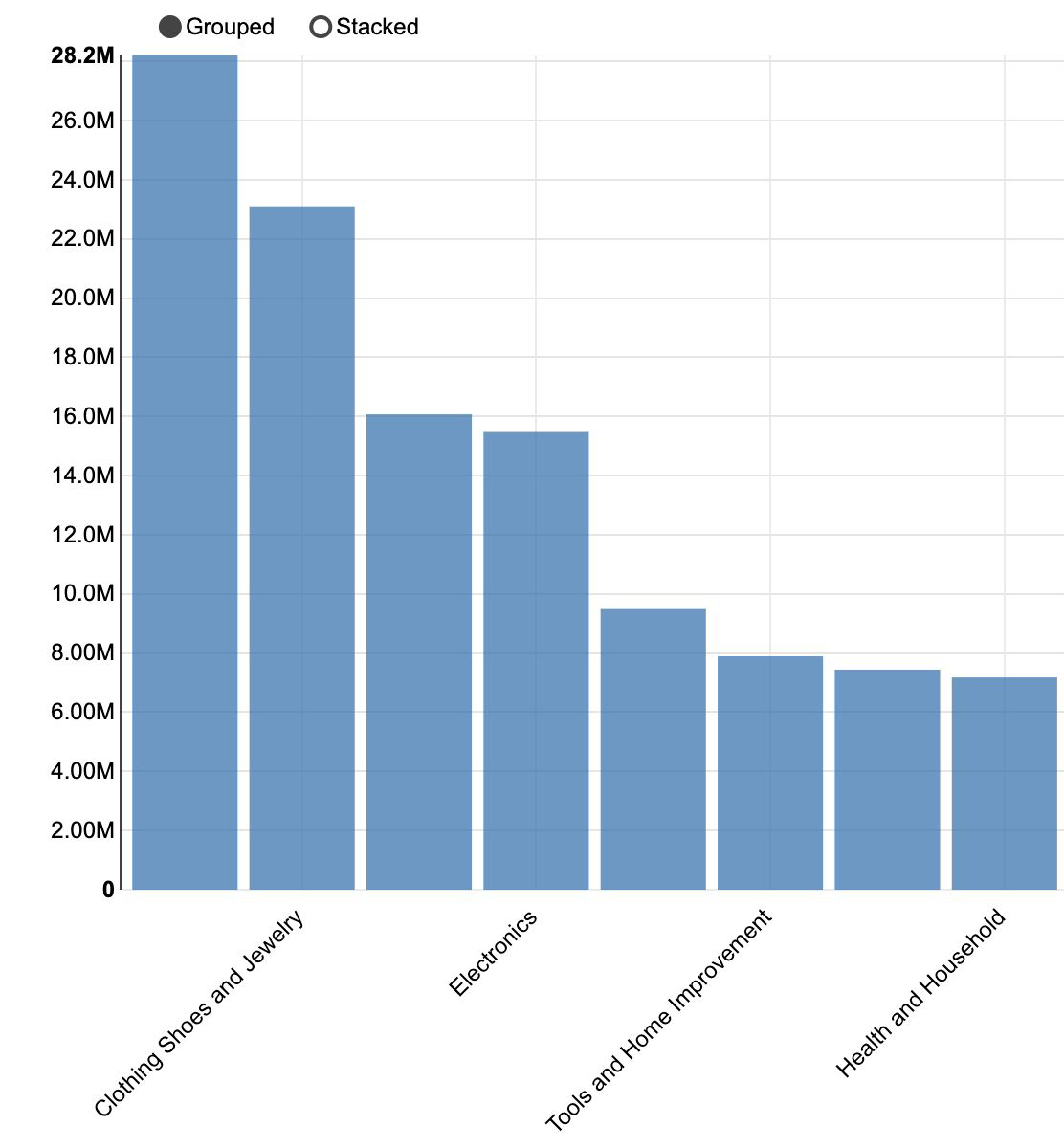}
    \caption{Category vs Ratings Count}
    \label{fig:enter-label}
\end{figure}

\subsubsection{Yearly Reviews Count}
Look at the total number of reviews uploaded every year, we find a sudden spike in yearly review counts in 2013 - indicating the year when Amazon became really popular as a site for e-commerce activities. We also notice that reviews have steadily increased upto 2019 after which it hit a plateau. This can mostly be attributed to the fact that almost all households now hold Amazon accounts.

The data in 2023 is incomplete which indicates the sudden drop in review counts.

\begin{figure}[h]
    \centering
    \includegraphics[width=1\linewidth]{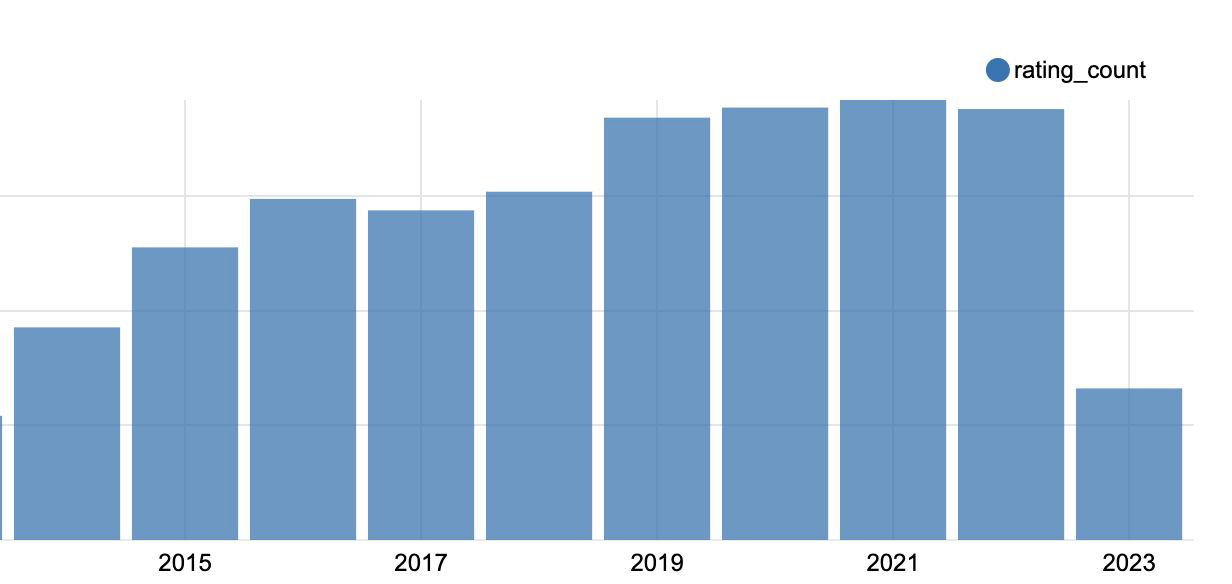}
    \caption{Year vs Ratings Count}
    \label{fig:enter-label}
\end{figure}

\section{Collaborative Analysis}

Once we individually analyzed every dataset, we attempted to collate our findings on the few fields common between the datasets. Here are some interesting findings from our analysis.

\subsection{Orders vs Days of the Week}

We count the total number of orders for each day of the week. Weeks are numbered from 0 to 6 with 0 representing Sunday and 6 representing Saturday.

\medskip
\noindent
In the \textbf{Instacart dataset} we calcuate summaries for all orders and orders that were reordered. Both show a similar trend with the \textbf{start of the week being especially busy} as households restock for the week. The counts \textbf{decrease during midweek} and start increasing again during the weekend.

\begin{figure}[h]
    \centering
    \includegraphics[width=\linewidth]{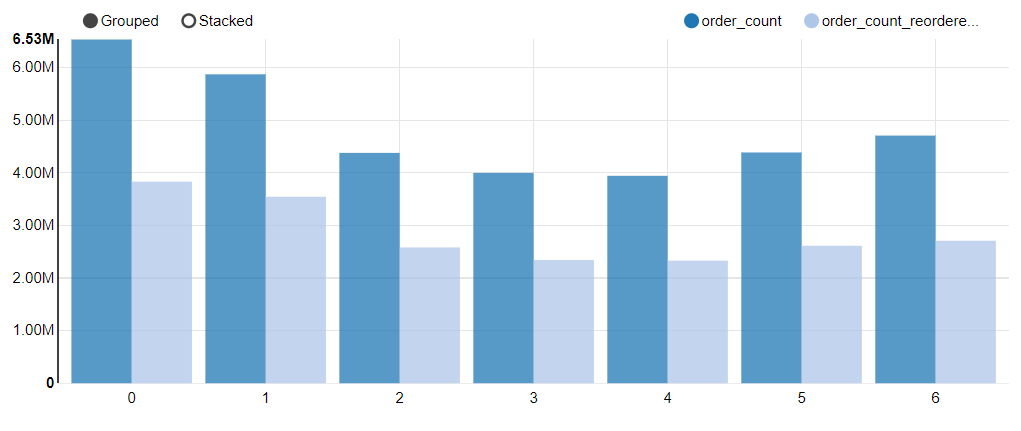}
    \caption{Instacart: Orders vs Day of the week}
    \label{fig:enter-label}
\end{figure}

The analysis holds true when we take a look at the \textbf{Percentage change in consumer spending} for \textbf{Sundays is higher} than any other day.

\begin{figure}[h]
    \centering
    \includegraphics[width=1\linewidth]{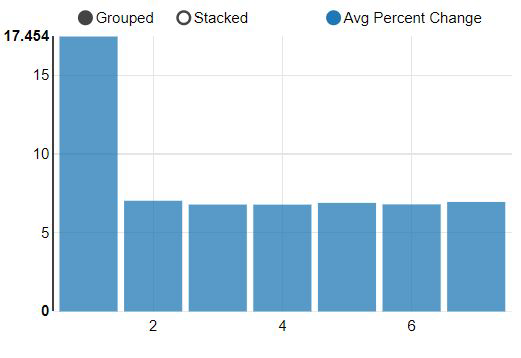}
    \caption{Percent Change: Average Change vs Day of the week}
    \label{fig:enter-label}
\end{figure}

Finally, we take a look at the \textbf{Amazon Reviews Dataset}. Here, we notice that the trend has switched up with a majority of the records getting created during midweek with the weekend counts being lower. This resonates with our previous hypothesis if we understand that the Amazon dataset looks at reviews rather than the actual orders. If we were to extrapolate these results, it would indicate that the orders were made during the weekend and reviews are put up during midweek.

\begin{figure}[h]
    \centering
    \includegraphics[width=1\linewidth]{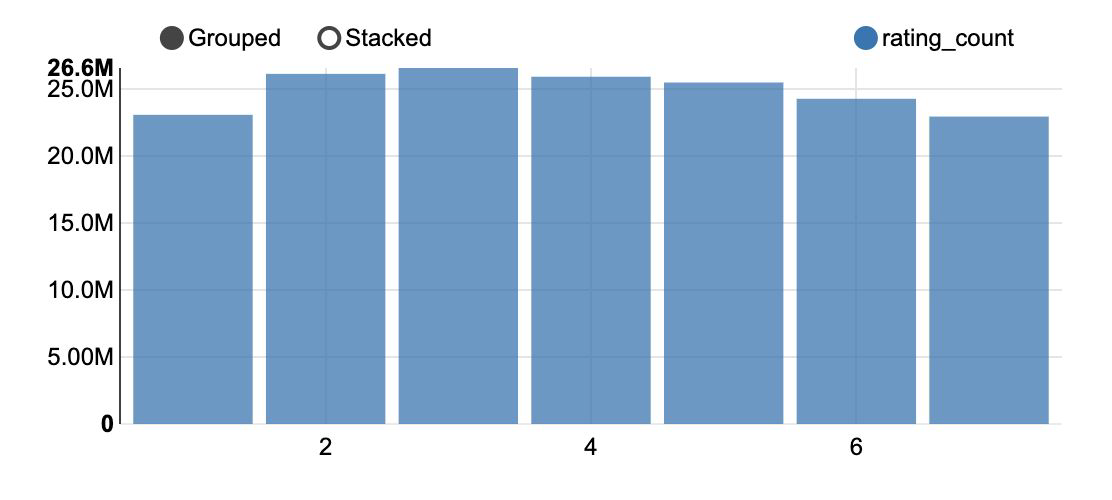}
    \caption{Amazon: Review Count vs Day of the Week}
    \label{fig:enter-label}
\end{figure}

\subsection{Orders vs Month of the Year}
If we were to perform a similar analysis with a different temporal domain, we would get further interesting insights.

\medskip
We can plot the distribution of the order count vs Month of the Week for the Amazon Dataset and the Walmart Dataset. For both these datasets we notice an interesting insight into the shopping habits of consumers.

\medskip
In both companies, we notice shopping and \textbf{order counts spike during the winter holiday months}. While in the case of Walmart, the peak is in January, for Amazon the orders start to peak from November and continue into the new year through January.

\begin{figure}[h!]
    \centering
    \includegraphics[width=1\linewidth]{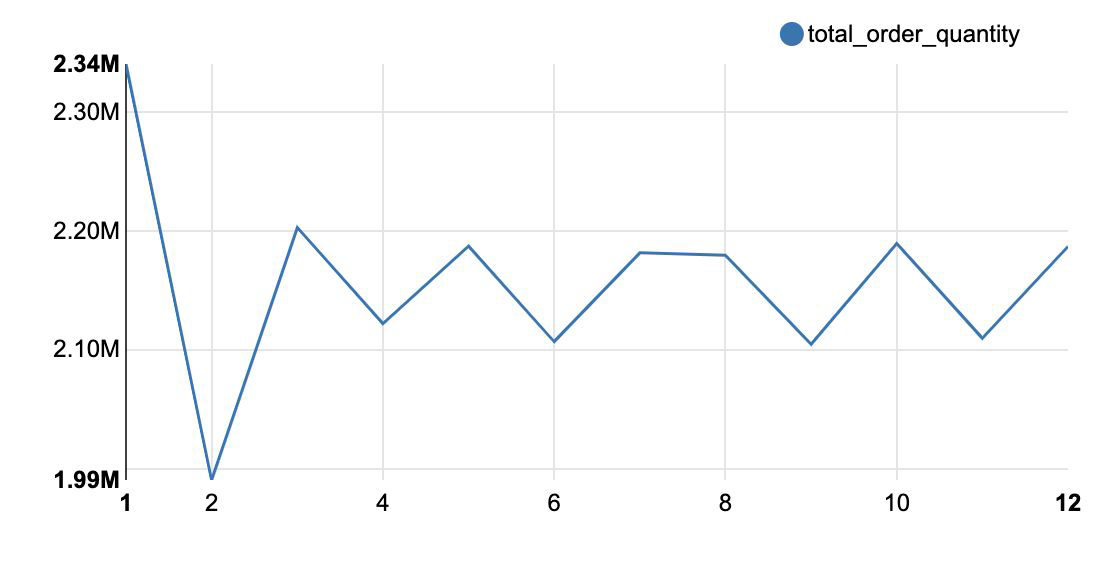}
    \caption{Walmart: Order counts vs Month of the Year}
    \label{fig:enter-label}
\end{figure}

\begin{figure}[h!]
    \centering
    \includegraphics[width=1\linewidth]{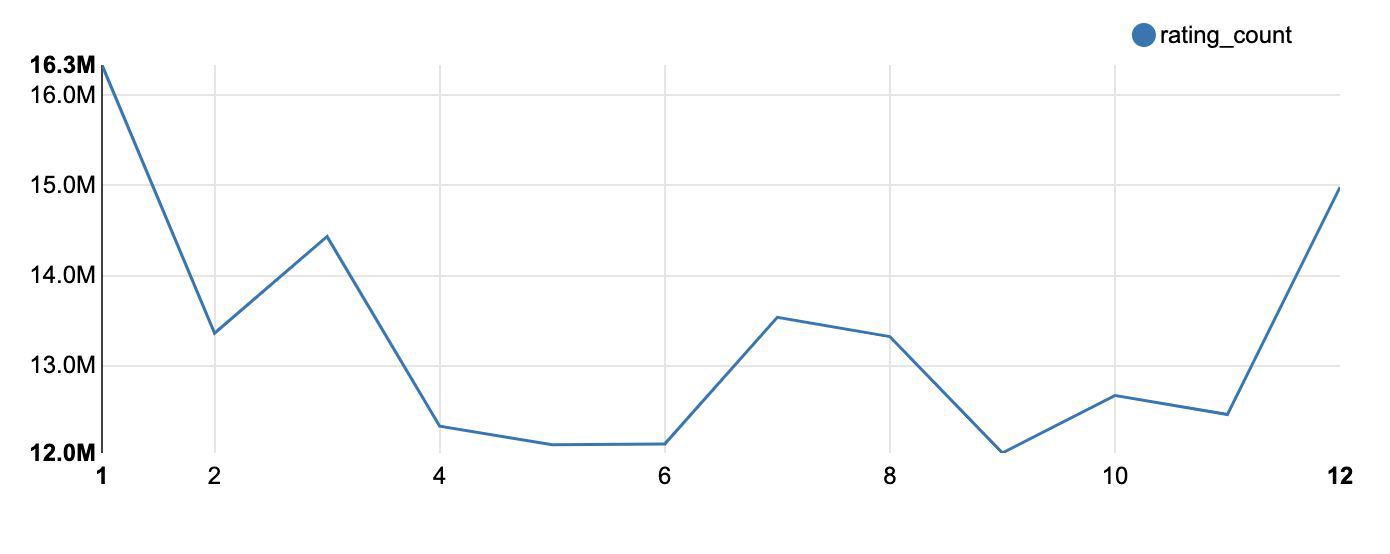}
    \caption{Amazon: Review Counts vs Month of the Year}
    \label{fig:enter-label}
\end{figure}

\newpage
\subsection{Correlation Between Reviews and Search Trends for Gift Cards}

The analysis explores the correlation between reviews for gift cards on Amazon and Google search trends for the keyword "Amazon Giftcard" over the last 5 years. 

\begin{figure}[h!]
    \centering
    \begin{subfigure}{0.45\textwidth}
        \centering
        \includegraphics[width=\linewidth]{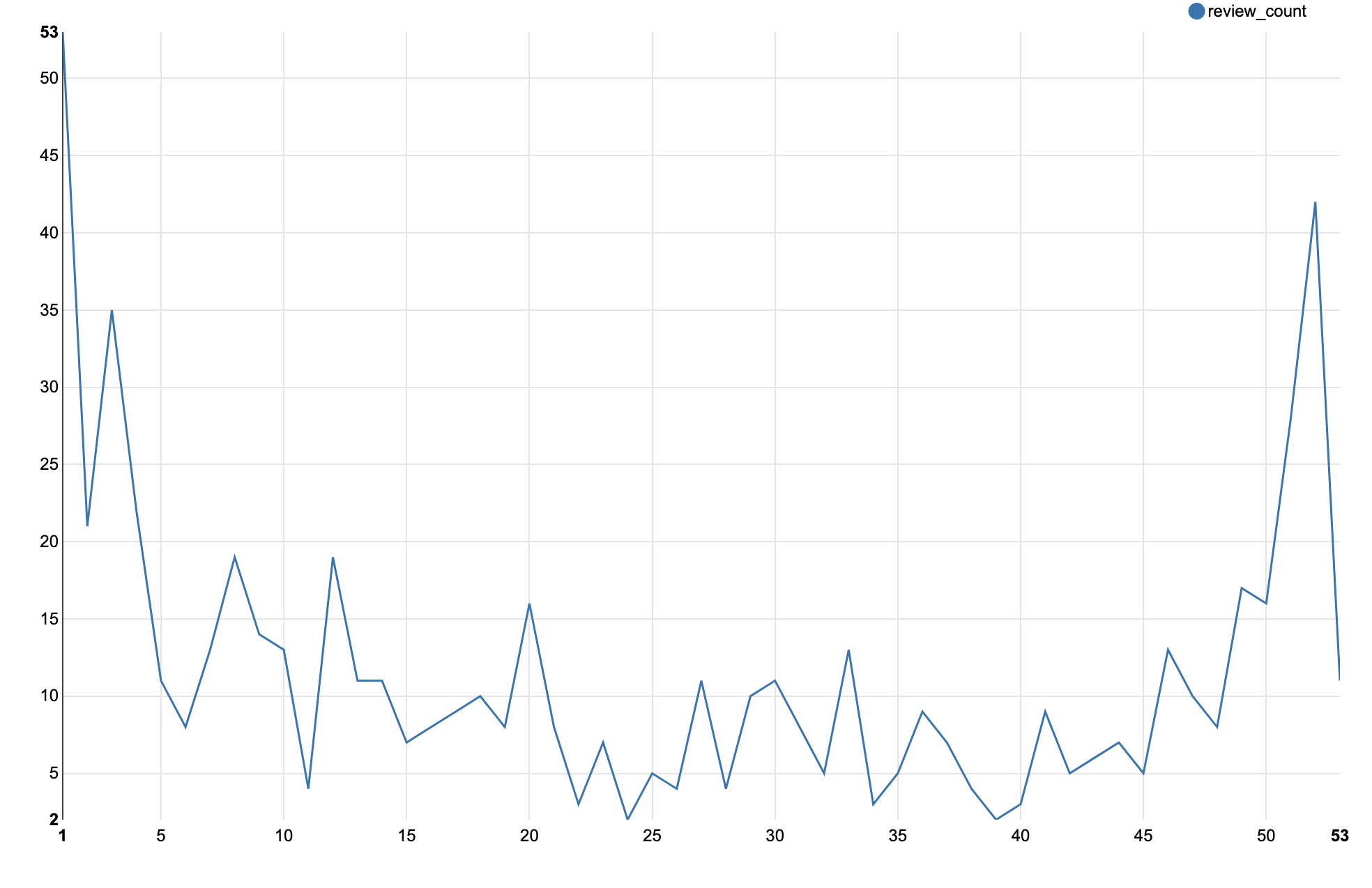} 
        \caption{Weekly reviews for Gift Cards on Amazon.}
        \label{fig:giftcard_reviews}
    \end{subfigure}
    \hfill
    \begin{subfigure}{0.45\textwidth}
        \centering
        \includegraphics[width=\linewidth]{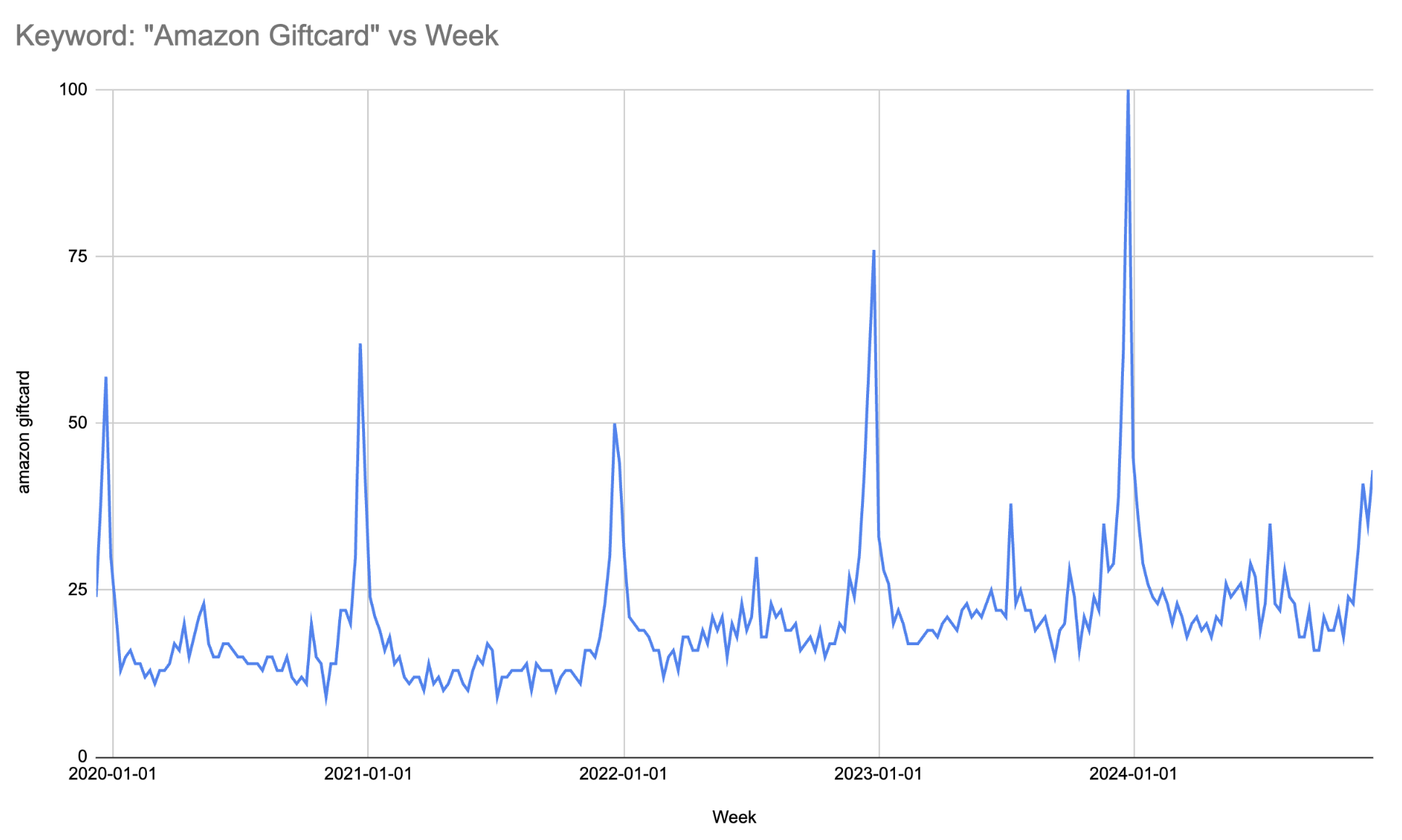} 
        \caption{Google search trends for "Amazon Giftcard."}
        \label{fig:google_trends}
    \end{subfigure}
    \caption{Correlation analysis between gift card reviews and Google search trends.}
    \label{fig:giftcard_analysis}
\end{figure}

The first plot (Figure \ref{fig:giftcard_reviews}) illustrates the average number of weekly reviews submitted for gift cards on Amazon. The data shows a significant spike in reviews during the last two weeks of December and the first two weeks of January. This pattern likely indicates that many customers leave reviews for gift cards after the winter holiday season.

The second plot (Figure \ref{fig:google_trends}) highlights weekly Google search trends for the keyword "Amazon Giftcard" from January 2020 onwards. It reveals a noticeable peak in searches during the winter holiday season, typically in December, suggesting a higher interest in purchasing gift cards during this period.

This combined analysis demonstrates a correlation between increased searches for gift cards during December and the subsequent reviews on Amazon, predominantly in late December and early January. The trend aligns with consumer behavior during the holiday season.

\section{Conclusions}

The results of this project underscore the transformative potential of big data analytics in optimizing retail operations. By integrating and analyzing datasets from diverse sources, such as Amazon reviews, Walmart sales, and Instacart orders, we uncovered actionable insights into consumer behavior, seasonal purchasing trends, and product preferences. These findings enable retailers to enhance inventory management, refine marketing strategies, and improve customer satisfaction.

A key highlight of this analysis is the identification of temporal patterns, such as the surge in gift card searches and reviews during the winter holiday season, and the midweek peak in Amazon review activity. Similarly, the start-of-week concentration of Instacart orders aligns with consumer restocking habits, showcasing how purchasing behaviors vary across platforms and timelines. The monthly analysis revealed consistent spikes in retail activity during the holiday months, particularly in December and January, reflecting seasonal consumer demand.

Furthermore, the correlation analysis between consumer actions and external factors, such as holidays and search trends, highlights opportunities for more targeted and efficient decision-making. By leveraging tools like Apache Spark for scalable data processing and visualization, this project showcases the importance of combining advanced analytics with domain knowledge to derive meaningful insights.

Ultimately, this work demonstrates the value of big data in fostering operational efficiency, improving customer experiences, and driving long-term growth in the retail sector. Future work could explore the integration of additional datasets, such as social media sentiment analysis, or the use of real-time analytics to capture emerging trends, enabling proactive and adaptive decision-making in a dynamic marketplace.

\section{Acknowledgement}
This work was made possible by downloading data from the Google Trends API, Kaggle, Amazon Datasets, Data.world and Data Gov.

Special thanks to the NYU Dataproc team to provide us with a distributed computing platform.

Last but not least, Professor Yang Tang for providing us with the skills to analyze Big datasets using Apache Spark.

\section{Repository Access}
The complete implementation for this project, including preprocessing, aggregation, analysis, and visualization code, is available on the GitHub repository:

\noindent\url{https://github.com/shrishriyesh/Retail-Market-Analysis}

The repository contains detailed notebooks, Scala scripts, and Zeppelin notebooks that demonstrate the methodologies and insights derived from the datasets. It serves as a resource for replicating the results and extending the analysis further.

\section*{References}

\begin{enumerate}
    \item Instacart Market Basket Analysis. Kaggle. (n.d.). \nolinkurl{https://www.kaggle.com/c/instacart-market-basket-analysis/data}.
    \item Amazon Reviews Dataset. (n.d.). Via Huggingface Datasets - Amazon Reviews’23. \nolinkurl{https://amazon-reviews-2023.github.io/dataloading/huggingface.html}.
    \item State of Connecticut - Percent Change in Consumer Spending. Data.gov. (2024, December 13). Available \href{https://catalog.data.gov/dataset/percent-change-in-consumer-spending-january-2020-through-the-present}{here}.

    \item Google. \emph{Google Trends Data}. Accessed December 13, 2024. \nolinkurl{https://trends.google.com/trends/}.
    \item Ahmed Mnif. \emph{Walmart Retail Dataset}. Data.world. Accessed December 13, 2024. \nolinkurl{https://data.world/ahmedmnif150/walmart-retail-dataset}.

\end{enumerate}

\end{document}